\documentclass[conference]{IEEEtran}
\ifCLASSINFOpdf
\usepackage[pdftex]{graphicx}
\else
\fi
\begin{document}
%
\title{Aspects of Sustainable Test Processes}

\author{\IEEEauthorblockN{Armin Beer}
\IEEEauthorblockA{Beer Test Consulting\\
Baden, Austria\\
info@arminbeer.at\\}
\and
\IEEEauthorblockN{Michael Felderer}
\IEEEauthorblockA{University of Innsbruck\\
Innsbruck, Austria\\
michael.felderer@uibk.ac.at}
\and
\IEEEauthorblockN{Tobias Lorey}
\IEEEauthorblockA{University of Innsbruck\\
Innsbruck, Austria\\
tobias.lorey@student.uibk.ac.at\\
}
\and
\IEEEauthorblockN{Stefan Mohacsi}
\IEEEauthorblockA{Atos IT Solutions and Services GmbH\\
Vienna, Austria\\
stefan.mohacsi@atos.net}}


%


\maketitle

\begin{abstract}
Testing is a core software development activity that has huge potential to make software development more sustainable. In this paper, we discuss how environmental, social, economic, and technical sustainability map onto the activities of test planning, design, execution, and evaluation.
\end{abstract}


%
\IEEEpeerreviewmaketitle

\section{Introduction}
Software development involves complex and dynamic systems. Changing technologies and markets in industry require improvements in software testing. To ensure a sustainable positive impact of improvement measures in a software project, multiple perspectives, including cultural, economic, and social issues, have to be taken into account  \cite{al2016engineering,penzenstadler2013generic,penzenstadler2013towards}. The sustainability definition of the Brundtland Commission \cite{dao2011green}, i.e., “meeting the needs of the present without compromising the ability of future generations to meet their own needs,” has environmental, social,  economic, and technical consequences that also impact how software testing should be conducted.

Experience, for instance in government projects, shows that software testing consumes resources and time comparable to the development tasks of the project \cite{mohacsi2019software}. An increase in complexity of requirements and architecture due to frequent adaptions in a project with a duration of 3-5 years is observed. Time pressure and cost restrictions prevent many projects from performing regenerative testing. The goal to perform software testing with \textit{“minimized environmental impact, a sufficient economic balance, and good working conditions”} \cite{penzenstadler2013towards} is not taken into account.

Currently, there is little guidance on how to integrate “green” strategies in the test process and how to align them to testing activities. Penzenstadler and Femmer \cite{penzenstadler2013generic} published a generic model for sustainability. However, they do not explicitly address the sustainability aspects of software testing. Testing activities require time and resources. Testing tasks consume energy and resources because large-scale projects have a long development and maintenance period. Server-client systems, e.g., for regression testing and performance testing, consume a lot of energy.

Experiences from large-scale projects \cite{mohacsi2019software} indicate the following problems:
\begin{itemize}
    \item Release and iteration planning neglects waste of resources and energy consumption.
    \item Lack of testability leads to increased complexity from version to version \cite{felderer2012estimating}.
    \item Knowledge networking and technology management are not taken into account as a long-term investment in future business.
\end{itemize}

As a consequence, the objective of setting up a sustainable test process is not taken into account \cite{dao2011green} .

\section{Dimensions of Sustainability in the Test Process}
The sustainability dimensions consist of the following viewpoints: \textit{environmental, social, economic, technical, and individual} sustainability \cite{penzenstadler2013generic}. They are depicted in Figure 1 with corresponding test process activities.

\begin{figure*}
\centering
\includegraphics[width=11cm]{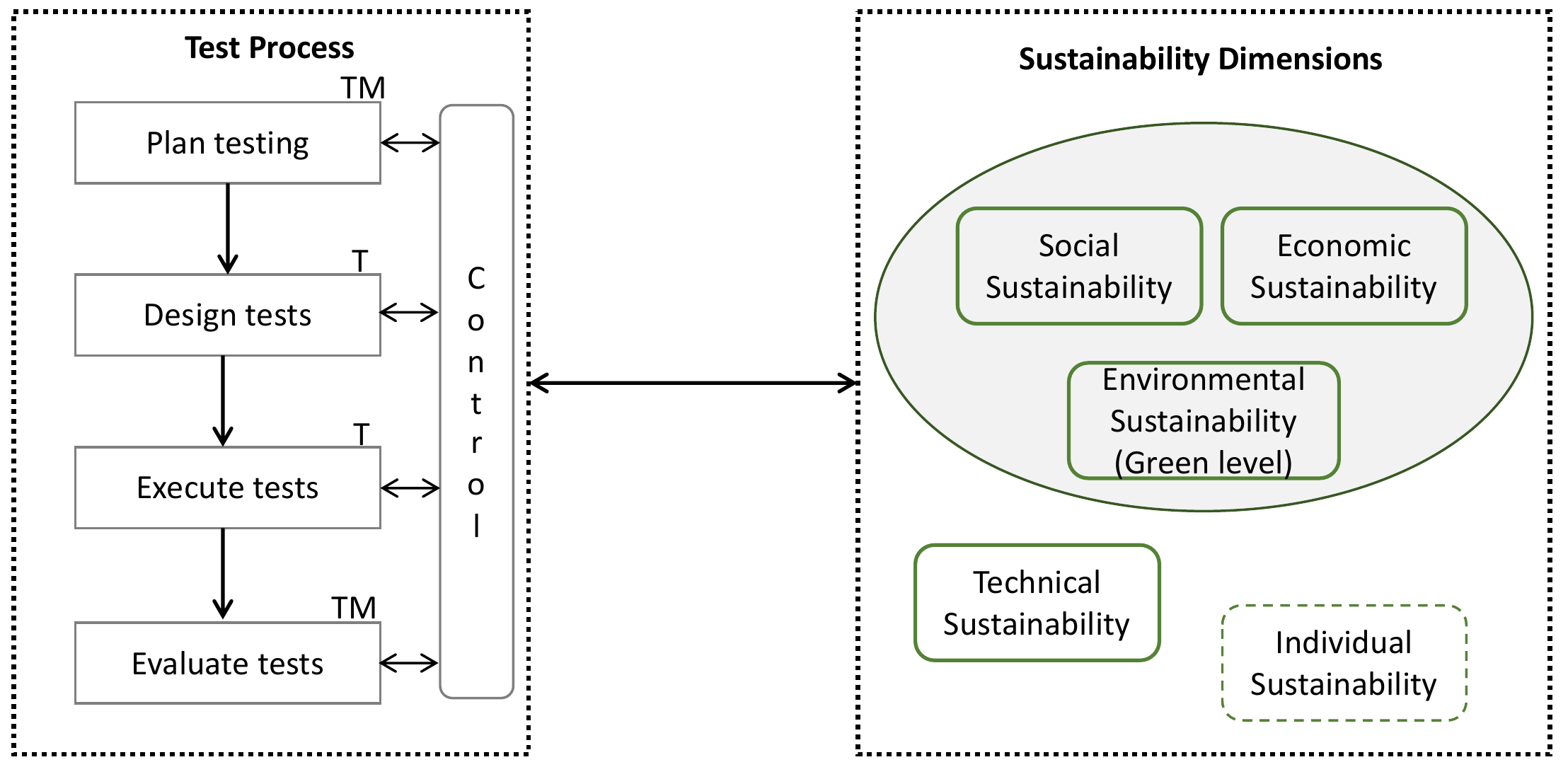}
\caption{Relationship between test process activities, the roles of a test manager (TM) and tester (T), and the sustainability dimensions \cite{penzenstadler2013generic}.}
\label{sustainable_testing-figure}
\end{figure*}

The \textit{environmental} dimension is a central “green” objective, for instance energy efficiency in software engineering \cite{calero2015green}.

The \textit{social} dimension aims at social capital as well as social communities and poses the question of how software impacts society in terms of social constructs and communication \cite{penzenstadler2013generic}. Further values of social sustainability include the health of employees, employment opportunities, the application of human rights at the workplace, and social networks within and outside an organization \cite{al2016engineering}. In the context of software testing, these principles apply to employees directly involved in the testing process, such as testers, test managers, and test engineers, as well as structures within an organization and society as a whole. Testing is also well-suited for onboarding new employees, and is therefore an activity that promotes  inclusiveness.

\textit{Economic} sustainability aims at reducing the costs of test execution and the maintenance of test artifacts. Automated regression tests are a typical measure for improving the economic sustainability of a software project. The additional effort for automating the tests will pay off after a certain number of test cycles and result in considerable savings in terms of manual effort and resources \cite{felderer2014industrial}. Moreover, every change to the system under test requires an adaptation of the corresponding test artifacts. Thus, the effort required to maintain test cases, test data, and scripts also constitutes a major economic factor in a typical software project. It can be significantly reduced by applying a model-based test (MBT) \cite{utting2012taxonomy,felderer2016model} approach. MBT relies on explicit models that encode information about the system under test. Test cases, test data, and scripts can be automatically generated from the model by an MBT tool.

\textit{Technical} sustainability is related, for instance, to long-living test environments for performance tests and test automation and their adequate evolution \cite{felderer2012estimating}. In this context, today’s fast-paced changes of the available technologies, tools, and interfaces can be a major challenge. For instance, a switch to a new user interface technology of the system under test can invalidate the existing test automation approach if the new technology is not supported by the test automation tool. Other examples include new network protocols, devices, operating systems, and so on. A possible mitigation strategy is to focus on maintainability and expandability aspects during the design of the test framework. Also, testability aspects should be taken into account when deciding about changes to the system under test. In addition, technology management and a collaboration model leverage knowledge networking in an organization if experts are willing to help each other.

The green level of sustainability aims at test efficiency and effectiveness of testing in general, thereby reducing resource and energy consumption. The following sustainability capabilities support these goals:\\
(1)	\textit{Plan testing:} Professional requirements management, reduction of complexity, and good testability are preconditions to save resources and reduce waste when performing test activities. Release decisions should focus on quality.\\ 
(2)	\textit{Design tests:} Test cases based on functional and non-functional requirements are designed by testers with domain know-how and technical skills. The degree of experience of testers influences the quality of the test cases as well as the cost of maintenance and execution of test cases.\\ 
(3)	\textit{Execute tests:} Test case execution is the most expensive task. Every corrected release has to be retested and regression-tested automatically or manually. The degree of testability has a strong impact on resource and time consumption.\\
(4)	\textit{Evaluate tests:} Long-term maintenance and use of systems require a measurement system to assess the trade-off, e.g., between technical sustainability criteria and economic criteria. Decisions of a test manager should be supported by a dashboard.

To introduce it into a software project, two distinct phases are recommended: 1) analysis and 2) application and assessment. The analysis phase consists of instantiating the generic sustainability model, and the application and assessment phase consists of specifying responsibilities and monitoring sustainability using previously defined indicators and metrics \cite{lago2015framing}.

\section{Conclusion}
Changing requirements, complexity, and a rigid time scale of projects are reasons why the long-term benefits of a sustainable test process are often not taken into account. We introduced dimensions and possible enhancements into the test process. However, concrete guidelines and measures are needed to promote environmental sustainability in software testing.

\bibliographystyle{IEEEtran}
\bibliography{references}

\begin{thebibliography}{10}
\providecommand{\url}[1]{#1}
\csname url@samestyle\endcsname
\providecommand{\newblock}{\relax}
\providecommand{\bibinfo}[2]{#2}
\providecommand{\BIBentrySTDinterwordspacing}{\spaceskip=0pt\relax}
\providecommand{\BIBentryALTinterwordstretchfactor}{4}
\providecommand{\BIBentryALTinterwordspacing}{\spaceskip=\fontdimen2\font plus
\BIBentryALTinterwordstretchfactor\fontdimen3\font minus
  \fontdimen4\font\relax}
\providecommand{\BIBforeignlanguage}[2]{{%
\expandafter\ifx\csname l@#1\endcsname\relax
\typeout{** WARNING: IEEEtran.bst: No hyphenation pattern has been}%
\typeout{** loaded for the language `#1'. Using the pattern for}%
\typeout{** the default language instead.}%
\else
\language=\csname l@#1\endcsname
\fi
#2}}
\providecommand{\BIBdecl}{\relax}
\BIBdecl

\bibitem{al2016engineering}
M.~Al~Hinai and R.~Chitchyan, ``Engineering requirements for social
  sustainability,'' in \emph{ICT for Sustainability 2016}.\hskip 1em plus 0.5em
  minus 0.4em\relax Atlantis Press, 2016.

\bibitem{penzenstadler2013generic}
B.~Penzenstadler and H.~Femmer, ``A generic model for sustainability with
  process-and product-specific instances,'' in \emph{Proceedings of the 2013
  workshop on Green in/by software engineering}, 2013, pp. 3--8.

\bibitem{penzenstadler2013towards}
B.~Penzenstadler, ``Towards a definition of sustainability in and for software
  engineering,'' in \emph{Proceedings of the 28th Annual ACM Symposium on
  Applied Computing}, 2013, pp. 1183--1185.

\bibitem{dao2011green}
V.~Dao, I.~Langella, and J.~Carbo, ``From green to sustainability: Information
  technology and an integrated sustainability framework,'' \emph{The Journal of
  Strategic Information Systems}, vol.~20, no.~1, pp. 63--79, 2011.

\bibitem{mohacsi2019software}
S.~Mohacsi and R.~Ramler, ``Why software testing fails: Common pitfalls
  observed in a critical smart metering project,'' in \emph{International
  Conference on Software Quality}.\hskip 1em plus 0.5em minus 0.4em\relax
  Springer, 2019, pp. 73--92.

\bibitem{felderer2012estimating}
M.~Felderer and A.~Beer, ``Estimating the return on investment of defect
  taxonomy supported system testing in industrial projects,'' in \emph{2012
  38th Euromicro Conference on Software Engineering and Advanced
  Applications}.\hskip 1em plus 0.5em minus 0.4em\relax IEEE, 2012, pp.
  426--430.

\bibitem{calero2015green}
C.~Calero and M.~Piattini, \emph{Green in software engineering}.\hskip 1em plus
  0.5em minus 0.4em\relax Springer, 2015, vol.~3.

\bibitem{felderer2014industrial}
M.~Felderer, A.~Beer, J.~Ho, and G.~Ruhe, ``Industrial evaluation of the impact
  of quality-driven release planning,'' in \emph{Proceedings of the 8th
  ACM/IEEE International Symposium on Empirical Software Engineering and
  Measurement}, 2014, pp. 1--8.

\bibitem{utting2012taxonomy}
M.~Utting, A.~Pretschner, and B.~Legeard, ``A taxonomy of model-based testing
  approaches,'' \emph{Software testing, verification and reliability}, vol.~22,
  no.~5, pp. 297--312, 2012.

\bibitem{felderer2016model}
M.~Felderer, P.~Zech, R.~Breu, M.~B{\"u}chler, and A.~Pretschner, ``Model-based
  security testing: a taxonomy and systematic classification,'' \emph{Software
  Testing, Verification and Reliability}, vol.~26, no.~2, pp. 119--148, 2016.

\bibitem{lago2015framing}
P.~Lago, S.~A. Ko{\c{c}}ak, I.~Crnkovic, and B.~Penzenstadler, ``Framing
  sustainability as a property of software quality,'' \emph{Communications of
  the ACM}, vol.~58, no.~10, pp. 70--78, 2015.

\end{thebibliography}
%

\end{document}